\documentclass[12pt,oneside,letterpaper]{article}
\pdfoutput=1
\usepackage{graphics}
\usepackage{amsmath, amssymb, amsthm}
\usepackage[pdftex]{graphicx}
\usepackage{hyperref}
\author{Rajendra K. Bera\footnote{Corresponding author; rbera@iiitb.ac.in.} {} and 
Vikram Menon}
\title{A new interpretation of superposition, entanglement, and measurement in 
quantum mechanics}
\date{}
\begin{document}
\maketitle

\section*{\textit{Abstract}}
We present a new interpretation of the terms superposition, entanglement, and 
measurement that appear in quantum mechanics. We hypothesize that the structure 
of the wave function for a quantum system at the sub-Planck scale has a 
deterministic cyclic structure. Each cycle comprises a sequential succession of 
the eigenstates that comprise a given wave function. Between unitary operations 
or measurements on the wave function, the sequential arrangement of the current 
eigenstates chosen by the system is immaterial, but once chosen it remains fixed 
until another unitary operation or measurement changes the wave function. The 
probabilistic aspect of quantum mechanics is interpreted by hypothesizing a 
measurement mechanism which acts instantaneously but the instant of measurement 
is chosen randomly by the classical measuring apparatus over a small but finite 
interval from the time the measurement apparatus is activated. At the instant the 
measurement is made, the wave function irrevocably collapses to a new state 
(erasing some of the past quantum information) and continues from thereon in that 
state till changed by a unitary operation or a new measurement.
\section*{Introduction}
Modern quantum mechanics is an axiomatic system and is based on the following 
four postulates $\cite{R1}$: 

\begin{description}
\item[\textbf{Postulate 1:}] Associated to any isolated physical system is a complex 
vector space with inner product (that is, a Hilbert space in finite space) 
known as the \textit{state space} of the system. The system is completely 
described by its \textit{state vector}, which is a unit vector in the system's 
state space.\hspace{1mm} $\blacksquare$

\item[\textbf{Postulate 2:}] The evolution of a closed quantum mechanical system is 
described by a \textit{unitary transformation}. That is, the state 
$|\psi(t_{1})\rangle$ of the system at time $t_{1}$ is related to the state 
$|\psi(t_{2})\rangle$ of the system at time $t_{2}$ by a unitary operator $U$ 
which depends only on the times $t_{1}$ and $t_{2}$,
\begin{equation*}
  |\psi(t_{2})\rangle = U|\psi(t_{1})\rangle. \hspace{2mm} \blacksquare 
\end{equation*}
A more refined version of Postulate 2 is the linear Schr\"{o}dinger 
equation, which describes the evolution of a closed quantum system in continuous 
time,
\begin{equation*}
 i\hbar\frac{d|\psi(t)\rangle}{dt} = H|\psi(t)\rangle. 
\end{equation*}
Here $\hbar$ is the reduced Planck's constant, and $H$ is a fixed Hermitian operator 
known as the quantum \textit{Hamiltonian} operator of the closed system. $H$ is 
a self-adjoint operator which acts on the state space and is related to the 
total energy of the system.

\item[\textbf{Postulate 3:}] Quantum measurements are described by a collection 
$\{M_{m}\}$ of \textit{measurement operators}. These are operators, which act on 
the state space of the system being measured. The index $m$ refers to the measurement 
outcomes that may occur in the experiment. If the state of the quantum system is 
$|\psi\rangle$ just before the measurement, then the probability that result $m$ 
occurs is given by
\begin{equation*}
 p(m) = \langle \psi|M^{\dagger}_{m}M_{m}|\psi\rangle, 
\end{equation*}
and the state of the system after the measurement is,
\begin{equation*}
 \frac{M_{m}|\psi\rangle}{\sqrt{\langle \psi|M^{\dagger}_{m}M_{m}|\psi\rangle}}, 
\end{equation*}
where the measurement operators satisfy the \textit{completeness condition},
\begin{equation*}
 \sum_{m}M^{\dagger}_{m}M_{m}=I. 
\end{equation*}
The completeness condition expresses the requirement that the respective 
probabilities associated with each state of $|\psi\rangle$ must sum to one: 
\begin{equation*}
 I = \sum_{m}{p(m)} = \sum_{m}{\langle \psi|M^{\dagger}_{m}M_{m}|\psi\rangle}. 
\hspace{2mm} \blacksquare 
\end{equation*}

\item[\textbf{Postulate 4:}] The state space of a composite physical system is the 
tensor product of the state spaces of the component physical systems. Moreover if 
we have systems numbered $1$ through $n$, and system number $i$ is prepared in 
the state $|\psi_{i}\rangle$ then the joint state of the total system is 
$|\psi_{1}\rangle \otimes |\psi_{2}\rangle \otimes \ldots \otimes 
|\psi_{n}\rangle.$ \hspace{1mm} $\blacksquare$
\end{description}

As is expected of axiomatic systems, the four postulates are independent of one 
another.

It is, of course, widely known that while quantum mechanics is the crown jewel 
of theoretical physics, it is non-intuitive and axiomatically different from 
classical physics. In quantum mechanics, while the wave function $|\psi\rangle$, 
introduced in postulate 2, is completely deterministic, its value, even using proxies, 
cannot be determined by a classical measurement system. A measurement made on a 
quantum system is governed by a separate postulate (postulate 3) and yields, in 
general, only partial non-deterministic information about the state of the system 
immediately prior to the measurement. It is not possible to know the complete 
state of an unknown quantum system by making multiple measurements on the same 
system or even on multiple available copies of the system. The very first 
measurement on a system irreversibly ``collapses'' the state of the system and 
it does that in a probabilistic manner. In fact, all the probabilities associated 
with quantum mechanics are enshrined only in postulate 3 and not in the other 
three postulates. The wave function of postulate 2 is an abstract mathematical 
entity; its origin and any underlying structure at the sub-Planck level that might 
support it are unknown and open to further speculation or interpretation. 
Postulate 3 does not tell us what the underlying measurement process is, it only 
tells us what the measurement result will be only in a probabilistic sense. So 
the measurement process too is open to further speculation or interpretation. 
Note that measurement is a nonlinear phenomenon and it is not governed by the 
Schr\"{o}dinger equation. Indeed, one is immediately struck by the 
contradictory nature of postulates 2 and 3. As David Albert notes $\cite{R2}$: 
\begin{quote}
The dynamics and the postulate of collapse are flatly in contradiction 
with one another ... the postulate of collapse seems to be right about what 
happens when we make measurements, and the dynamics seems to be bizarrely 
\emph{wrong} about what happens when we make measurements, and yet the 
dynamics seems to be \emph{right} about what happens whenever we 
\emph{aren't} making measurements.
\end{quote}

Lalo\"{e} neatly summarizes the dilemmas produced by the two postulates $\cite{R3}$:
\begin{quote}
Obviously, having two different postulates for the evolution of the same 
mathematical object is unusual in physics; the notion was a complete novelty when 
it was introduced, and still remains unique in physics, as well as the source of 
difficulties. Why are two separate postulates necessary? Where exactly does the 
range of application of the first stop in favor of the second? More precisely, 
among all the interactions - or perturbations - that a physical system can 
undergo, which ones should be considered as normal (Schr\"{o}dinger 
evolution), which ones as a measurement (wave packet reduction)? Logically, we 
are faced with a problem that did not exist before [in classical physics], when 
nobody thought that measurements should be treated as special processes in 
physics.
\end{quote}

Superposition and entanglement of quantum states are two other intriguing aspects 
of quantum mechanics. Superposition of states which results from postulate 2 is 
the same as is generally understood in the mathematics of linear systems and in 
classical physics. Postulate 2 does not require one to view superposition where 
an electron is concurrently in spin-up and spin-down states. It is postulate 3 
related to measurement which allows us to adopt the implausible view that an 
electron can be concurrently in spin-up and spin-down states rather than in some 
in-between state. Entanglement, viewed at one time as spooky action at a distance, 
comes from postulates 1, 2 and 4. It is no longer viewed with suspicion $\cite{R4}$ 
since entangled particles are now routinely produced in experiments. John 
Bell $\cite{R5}$ showed hidden variables are not needed to explain the phenomenon 
and Aspect, \textit{et al} $\cite{R6}$ provided experimental verification that 
entanglement is real. We have already noted that a quantum system can exist in a 
continuum of states until it is measured. But this is not the only bizarre 
consequence of a measurement. When two or more particles are entangled then a 
measurement on any one particle or a combined measurement on a subgroup of 
particles will cause a `jump' to occur instantly on the remaining particles even 
if some or all of them are light years apart. Indeed, distance is irrelevant for 
entangled particles. A group of entangled particles have a \emph{distributed 
existence} in the sense that the group behaves as a single entity even when spread 
out in space. Such bizarre behaviors have been extensively verified in experiments.

The postulates of quantum mechanics do not enlighten us as to the nature of the 
underlying structure of the wave function at the sub-Planck scale that would 
support entanglement, and which, in turn, would lead to a wave function that 
follows the postulates of quantum mechanics. So, once again, the underlying 
structure of the wave function is open to speculation or interpretation. The 
interpretation does not predict results; that is the job of the 
Schr\"{o}dinger wave equation and the measurement postulate. However, for 
any interpretation to be acceptable it must be compatible with measured results; 
that is, it must put the mathematical model into correspondence with experience. 
The distinction between an axiomatic system and its interpretation has been 
elucidated by Hofstadter $\cite{R7}$ with great clarity in his book 
G\"{o}del, Escher, Bach. A given axiomatic system may have more than one 
interpretation.

\section*{Some earlier interpretations of quantum mechanics}
While the formalism of quantum mechanics is widely accepted, there is no single 
interpretation of it that is agreeable to everyone. The disagreements essentially 
stem from the incompatibility that exists between postulates 2 and 3. Indeed, 
without postulate 3 telling us what we can observe, the equations of quantum 
mechanics would be just pure mathematics that would have no physical meaning at 
all. Note also that any interpretation can come only \emph{after} an 
investigation of the logical structure of the postulates of quantum mechanics is 
made. We briefly digress to elaborate what we mean by an interpretation in the 
context of this paper.

Newtonian mechanics does not define the structure of matter. How we interpret or 
model the structure of matter is largely an issue separate from Newtonian 
mechanics. However, any model of the structure of matter we propose is expected 
to be such that it is compatible with Newton's laws of motion in the realm where 
Newtonian mechanics rules. If it is not, then Newtonian mechanics, as we know it, 
would have to be abandoned or modified, or the model of the structure of matter 
would have to be abandoned or modified. One may also have a partial 
interpretation and leave the rest in abeyance till further insight strikes us and 
leads us to a complete or a new interpretation. As we know, our understanding 
(interpretation) of the structure of matter has undergone several changes 
(including our current understanding of the subatomic structure of matter as 
enshrined in the still evolving standard model of particle physics) without 
affecting Newton's laws of motion. A question such as whether a particular result 
deduced from Newton's laws of motion is deducible from a given model of material 
structure is therefore not relevant.

Likewise, as long as our interpretation (or model) of superposition, entanglement, 
and measurement does not require the axioms of quantum mechanics to be altered, 
and as far as we can determine, it does not, none of the predictions made by 
quantum mechanics should be incompatible with our interpretation. This assertion 
is important because we make no comments on the Hamiltonian, which captures the 
detailed dynamics of a quantum system. Quantum mechanics does not tell us how to 
construct the Hamiltonian. In fact, real life problems seeking solutions in 
quantum mechanics need to be addressed in detail by physical theories built 
within the framework of quantum mechanics. The postulates of quantum mechanics 
provide only the scaffolding around which detailed physical theories are to be 
built.

Our interpretation and the postulates of quantum mechanics are two different but 
related things. Quantum mechanics leaves room for interpretation because the wave 
function is an abstract mathematical object. Neither its origin nor its 
underlying structure has been disclosed in the postulates of quantum mechanics. 
Furthermore, the mechanisms for superposition, entanglement, and measurement too 
have not been elucidated in quantum mechanics. Hence, as noted earlier, they too 
are open to interpretation. We have, in a sense, tried to provide a sub-Planck 
scale view of the wave function, superposition, entanglement, and measurement 
without affecting the postulates of quantum mechanics. The sub-Planck scale is 
chosen to provide us with the freedom to construct mechanisms for our 
interpretation that are not necessarily bound by the laws of quantum mechanics. 
In particular our interpretation does not have to satisfy the 
Schr\"{o}dinger wave equation because quantum mechanics is not expected to 
rule in the sub-Planck scale. The high point of our interpretation is that it is 
able to explain the measurement postulate as the inability of a classical 
measuring device to measure at a precisely predefined time.

In passing we note that there already exist several interpretations of quantum 
mechanics, each with its own bizarre elements. Human experiences of the world 
are entirely based on macroscopic objects which behave according to the laws of 
classical physics. Each attempted interpretation of quantum mechanics has 
therefore been in a language that lacks the appropriate concepts. Perhaps the 
most widely preferred by physicists is the Copenhagen interpretation, possibly 
followed by Everett's many world interpretation (a variant of which is favored 
by David Deutsch), and Bohm's pilot wave interpretation (favorably commented 
upon by John Bell). Each known interpretation contains concepts or elements that 
are alien to our everyday experience. No one knows what the reality of a quantum 
system could be.

\subsection*{\textit{The Copenhagen interpretation}} 
We are not aware of any systematic rendition of the Copenhagen interpretation in 
the scientific literature. As Jammer notes: ``The Copenhagen view is not a single, 
clear-cut, unambiguously defined set of ideas but rather a common denominator for 
a variety of related viewpoints'' $\cite{R8}$. The Copenhagen interpretation 
(also known as the `shut up and calculate' interpretation) was provided by Niels 
Bohr and Werner Heisenberg around 1927. In doing so, they extended the 
probabilistic interpretation of the wave function proposed by Max Born $\cite{R9}$.
Some interesting features included in the Copenhagen interpretation are: the 
position of a particle is essentially meaningless; the act of measurement causes 
an instantaneous collapse of the wave function and the collapsed state is randomly 
picked to be one of the many possibilities allowed for by the system's wave 
function; the fundamental objects handled by the equations of quantum mechanics 
are not actual particles that have an extrinsic reality but ``probability waves'' 
that merely have the capability of becoming real when an observer makes a 
measurement. Even then it does not explain entanglement; this experimentally 
verified ``nonlocality'' is a mathematical consequence of quantum theory $\cite{R10}$. 
In the Copenhagen interpretation one cannot describe a quantum system independently 
of a measuring apparatus. Indeed, it is meaningless to ask about the state of the 
system in the absence of a measuring system. For example, the Copenhagen 
interpretation not only does not explain how an electron goes through the two 
slits in the double slit experiment, it categorically states that even to ask such 
a question is meaningless and that we should restrict our comments to the observed 
interference pattern on the screen $\cite{R11}$. The role of the observer is 
central since it is the observer who decides what he wants to measure $\cite{R11}$.
Nevertheless, there is a clear demarcation between the quantum system being 
measured and the macroscopic measuring device (described by classical mechanics).

In Bohr's view $\cite{R11}$,
\begin{quote}
\dots{}  there is no quantum world. There is only abstract quantum physical 
description. It is wrong to think that the task of physics is to find out how 
nature is. Physics [only] concerns what we can say about nature.
\end{quote} 

This view is very different from that of Einstein's who believed that the job of 
physical theories is to `approximate as closely as possible to the truth of 
physical reality' $\cite{R11}$. 

\subsection*{\textit{Everett's many world interpretation}}
Hugh Everett III, in his doctoral dissertation of 1956 (see Ref. $\cite{R12}$ 
for the journal version), proposed what he called the ``relative state 
interpretation'', which is now generally known as the many world interpretation 
(the name was coined by Bryce DeWitt in the late 1960s). This interpretation is 
perhaps the most bizarre and yet perhaps the simplest (it is free of the 
measurement problem because Everett omits the measurement postulate) if one is 
willing to accept that we inhabit one of an infinite number of parallel worlds! 
Everett assumes that when a quantum system is faced with a choice such as a 
photon going through one or the other slit in the two-slit experiment, rather 
than the wave function entering a superposition, the entire world along with it 
splits into a number of worlds equal to the number of options available. These 
different worlds are identical to each other except for the different option 
chosen by the quantum system (such as the photon passing through the upper slit 
in one world and through the lower slit in another world). Until decoherence, 
that is, (spontaneous) interactions between a quantum system and its environment 
leads to the suppression of wave interference, sets in, the worlds overlap only 
in regions where interference is taking place. Decoherence causes the worlds to 
separate into non-interacting independent worlds. This allows Everett to avoid 
the non-intuitive problems related to measurement since there is no measurement 
process involved. When a world splits, observers in it will also split with it. 
Thus there will be other copies of the observers in parallel worlds, each of 
whom will see the specific outcome that appears in his respective split 
world $\cite{R11}$.

Physicists at the time ridiculed Everett's interpretation (he got his PhD alright 
and the work was published in Ref. $\cite{R12}$). It was said that ``Bohr gave 
him the brush-off when Everett visited him in Copenhagen'' $\cite{R11}$. 
He was so discouraged by the ridicule that he left physics, became a defense 
analyst, then a private contractor to the U.S. defense industry which made him 
a multimillionaire $\cite{R11,R13}$. Variants of the many-world interpretation 
are the multiverse interpretation, the many-histories interpretation, and the 
many-minds interpretation. 

\subsection*{\textit{Bohm's interpretation}}
In Bohm's interpretation $\cite{R14}$, which appeared in 1952 and predates 
Everett's interpretation, the whole universe is entangled and one cannot isolate 
one part of the universe from the other. Rather than interpret entanglement as 
some mysterious phenomenon, as in the Copenhagen interpretation, Bohm favors 
an interpretation which makes non-locality explicit. In his view, the interaction 
between entangled particles is not mediated by any conventional field known to 
physics (such as the electromagnetic field), but by a very special 
anti-relativistic \emph{quantum information field} (pilot wave) that does not 
diminish with distance and that binds the whole universe together. This field is 
an all pervasive field that is instantaneous. This field is not physically 
measurable but manifests itself in terms of non-local correlations. The idea is 
not only interesting but entirely derivable from the Schr\"{o}dinger equation. 
Consequently, in Bohm's interpretation, for example, the electron is a particle 
with well-defined position and momentum at any instant. However, the path an 
electron follows is guided by the interaction of its own pilot wave with the 
pilot waves of other entities in the universe. A major supporter of Bohm's 
interpretation was John Bell.

There are other interpretations which we omit from our discussion since a 
thorough review of them is not our intent. The widely different views from which 
the various interpretations follow are rather remarkable. In short, as yet there 
is no unique and fully satisfactory interpretation $\cite{R11}$.

\section*{A new interpretation}
Following the principle of Occam's razor that ``entities should not be multiplied 
unnecessarily'' or the law of parsimony, our interpretation adopts the viewpoint 
that the sub-Planck scale structure of the wave function is such that the wave 
function is in only one state at any instant but oscillates between its various 
``superposed'' component eigenstates. (There is no expenditure of energy in 
maintaining the oscillations.) To this we add a probabilistic measurement model 
which determines only the instantaneous eigenstate of the system at the instant 
of measurement. We essentially hypothesize a measurement mechanism which acts 
instantaneously but the instant of measurement is chosen randomly by the 
classical measurement apparatus over a small but finite interval from the time 
the measurement apparatus is activated. In particular, we regard measurement as 
the joint product of the quantum system and the macroscopic classical measuring 
apparatus. We do not explain how the collapse of the wave function occurs when a 
measurement is made. However, once a measurement is made, the wave function 
assumes the collapsed state. The notion of the collapse of the wave function upon 
measurement was introduced by Heisenberg in 1929 $\cite{R11}$.

\begin{figure}[htp]
\begin{center}
\includegraphics[width=60mm,height=35mm]{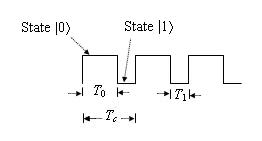}
\caption{A single particle system in superposition state}\label{fig:fig1}
\end{center}
\end{figure}

In our interpretation, superposed states appear as time-sliced in a cyclic manner 
such that the time spent by an eigenstate in a cycle is related to the complex 
amplitudes appearing in the wave function. Entangled states binding two or more 
particles appear in our interpretation as the synchronization of the sub-Planck 
level oscillation of the participating particles. Unlike the Copenhagen 
interpretation, in our interpretation it is not meaningless to ask about the 
state of the system in the absence of a measuring system. In essence, we seek an 
interpretation that can be related to the macroscopic world we live in and hence 
appears more intuitive to the human mind. We note that no rational interpretation 
of quantum mechanics in terms of images and concepts familiar to us from everyday 
experience in the macroscopic world has yet been found. Our's is perhaps the 
closest yet. The basic sub-Planck model that underpins our interpretation is as 
follows. Consider a qubit (quantum bit) described by the wave function
\begin{equation*}
    |\psi\rangle = \alpha|0\rangle + \beta |1\rangle
\end{equation*}
where the particle is in a superposed state comprising the eigenstates $|0\rangle$ 
and $|1\rangle$, and $\alpha$, $\beta$ are normalized complex constants such that 
$|\alpha|^{2} + |\beta^{2}| = 1$. The structure of the wave function $|\psi\rangle$, 
outside the realm of quantum mechanics and in the realm of sub-Planck scale that 
we now propose, is illustrated in Figure~\ref{fig:fig1}. 
Here we have two eigenstates, $|0\rangle$ and $|1\rangle$, and in a cycle time of 
$T_{c}$ the state of the particle oscillates between states $|0\rangle$ and 
$|1\rangle$. We assume $T_{c}$ to be much smaller than Planck time ($\ll 10^{-43}$ sec) 
to allow us to interpret the wave function independently of the Schr\"{o}dinger 
equation. (The implicit assumption here is that in some averaged sense, perhaps 
with some additional information, these oscillations will represent the wave 
function $|\psi\rangle$, say, analogous to the case of a volume of gas in 
classical mechanics, where random molecular motions, appropriately averaged, 
represent classical pressure, temperature, and density of a volume of gas.) It 
is not necessary for us to know the value of $T_{c}$. We only assert that it is 
a universal constant. Within a cycle, the time spent by the particle in state 
$|0\rangle$ is $T_{0} = |\alpha|^{2} T_{c}$ and in state $|1\rangle$ is 
$T_{1} = |\beta|^{2} T_{c}$ so that $T_{c} = T_{0} + T_{1}$. Superposition is 
interpreted here as the deterministic linear sequential progression of the 
particle's states $|0\rangle$ and $|1\rangle$. A measurement made on this 
particle will return the instantaneous state the particle was in at the instant 
of the measurement.

Our model of the measurement device is as follows. Let $\Delta t_{m}$ be the time 
interval during which some chosen measuring device makes a measurement. Here 
$\Delta t_{m}$ is assumed to be orders of magnitude greater than Planck time; otherwise 
its actual value is immaterial. Effectively, the device is assumed to make the 
measurement instantaneously at some instant, randomly chosen by the measuring 
device in the interval $\Delta t_{m}$. To avoid bias, we assume that the device can 
choose any instant in the interval $\Delta t_{m}$ with equal probability. Thus the 
source of indeterminism built into quantum mechanics via postulate 3 is 
interpreted here as occurring due to the classical measuring device's inability 
to measure at a precisely predefined time. As soon as the measurement is made, 
the quantum system assumes the measured state ``instantaneously'' as dictated by 
postulate 3.

\begin{figure}[htp]
\begin{center}
\includegraphics[width=110mm,height=70mm]{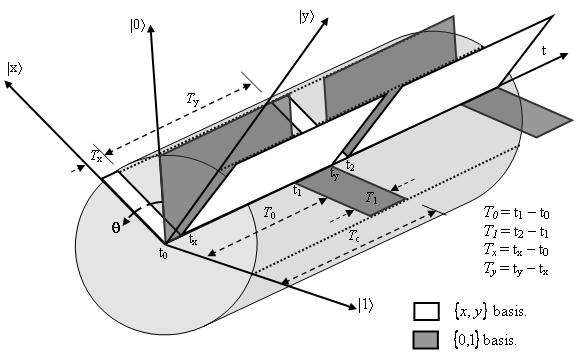}
\caption{Projection of a single particle system to $|0\rangle, |1\rangle$ and 
         $|x\rangle, |y\rangle$ bases.}\label{fig:fig2}
\end{center}
\end{figure}

Since the measurement device need not operate with the $\{|0\rangle, |1\rangle\}$ 
basis, we need a rule that allows us to go from one basis to another. The rule 
turns out to be rather simple. Let the measurement basis be $\{|x\rangle, 
|y\rangle\}$ obtained by rotating the basis $\{|0\rangle, |1\rangle\}$ anti-clockwise 
by the angle $\theta$, then $|0\rangle = \cos \theta|x\rangle + \sin \theta |y\rangle$ 
and $|1\rangle = \cos \theta |y\rangle - \sin \theta |x\rangle$ and conversely 
$|x\rangle = \cos \theta |0\rangle - \sin \theta |1\rangle$ and 
$|y\rangle = \sin \theta |0\rangle + \cos \theta |1\rangle$. In 
essence, Figure ~\ref{fig:fig2} shows how the vectors $|0\rangle, |1\rangle$ will 
be observed by a measuring device in the basis $\{|x\rangle, |y\rangle\}$ and the 
probabilities with which it will measure $|x\rangle$ or $|y\rangle$. Choosing a 
basis different from $\{|0\rangle, |1\rangle\}$ means changing the values of 
$t_{1}$ and $t_{2}$ to $t_{x}$ and $t_{y}$ and correspondingly re-labeling the 
eigenstates to $|x\rangle$ and $|y\rangle$.   
Thus, we can easily verify that 
\begin{align*}
  |\psi\rangle &= \alpha |0\rangle + \beta |1\rangle \\
               &= \alpha(\cos \theta|x\rangle + \sin \theta |y\rangle) +
                  \beta (\cos \theta |y\rangle - \sin \theta |x\rangle) \\
               &= (\alpha \cos \theta - \beta \sin \theta)|x\rangle + 
                  (\alpha \sin \theta + \beta \cos \theta)|y\rangle.
\end{align*}
Further, $T_{x}$ and $T_{y}$ corresponding to the time durations the system will 
be in state $|x\rangle$ and $|y\rangle$, respectively, with respect to $T_{c}$ in 
$\{|0\rangle, |1\rangle\}$ basis is given by
\begin{equation*}
    T_{x} = |\alpha \cos \theta - \beta \sin \theta|^{2} T_{c}, \hspace{2mm}
    T_{y} = |\alpha \sin \theta + \beta \cos \theta|^{2} T_{c},
\end{equation*}
\begin{equation*}
    T_{c} = T_{0}/|\alpha|^{2} = T_{1}/|\beta|^{2}.
\end{equation*}
Thus, if the measurement basis is $\{|x\rangle, |y\rangle\}$, the system, when 
measured, will randomly collapse to $|x\rangle$ or $|y\rangle$ with probability 
$T_{x}/T_{c} = |\alpha \cos \theta - \beta \sin \theta |^{2}$ or 
$T_{y}/T_{c} = |\alpha \sin \theta + \beta \cos \theta |^{2}$, respectively.

\begin{figure}[htp]
\begin{center}
\includegraphics[width=125mm,height=45mm]{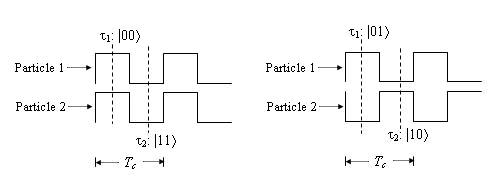}
\caption{Two-particle entangled systems; $|\psi_{1}\rangle$ (left), 
         $|\psi_{2}\rangle$ (right).}\label{fig:fig3}
\end{center}
\end{figure}

Finally, our model of entanglement requires that any unitary operation that 
causes entanglement, say, between two particles, also synchronizes their sub-
Planck level oscillations. This is shown in Figure ~\ref{fig:fig3} for the 
two-particle system Bell states,     
\begin{equation*}
 |\psi_{1}\rangle = (|00\rangle \pm |11\rangle)/\sqrt{2},
\end{equation*}
\begin{equation*}
 |\psi_{2}\rangle = (|01\rangle \pm |10\rangle)/\sqrt{2}.
\end{equation*}

Note that while entanglement results in a synchronous state for the two particles, 
the converse is not necessarily true. When a measurement is made on one of the 
entangled particles, both will collapse simultaneously. According to our model, 
the particles will collapse to the state they are in at the instant of measurement 
(such as $|\psi_{1}\rangle$ or $|\psi_{2}\rangle$ in Figure ~\ref{fig:fig3}), 
which is in accord with the measurement postulate (postulate 3). We do not know 
how Nature might accomplish the required synchronization.

It is, of course, clear that our interpretation cannot violate the uncertainty 
principle since the latest measurement on a system collapses the system according 
to the measurement postulate. Thus there can be no direct correlation between any 
earlier results of measurement on the system, and the succeeding measurement. 

Our interpretation, in essence, is yet one more underlying theory in which the 
nature and consistency of quantum theory can be investigated and clarified; it 
rests entirely on the notion of external observations because without it we have 
no means to ascribe a physical interpretation.


\section*{Application of the basic model}
We now provide a few examples of quantum systems to show that our interpretation 
is consistent with the outcomes of measurements made on those systems at any 
instant.

\subsection*{Measurement of a two-particle entangled system}
Recall that a measurement made on either particle in an entangled pair will 
automatically and instantaneously alter the state of the other particle. We are 
now confronted with two measurement possibilities: (1) measurement using 
commutating observables (such as of electron spin along the same axis); and 
(2) measurement using non-commutating observables (such as of electron spins 
along different axes).

(1) \textit{Commutating observables}. Consider an entangled pair of electrons, 
where $|0\rangle$ and $|1\rangle$, represent spin-up and spin-down, respectively. 
Then according to our model, if a measurement is made along the spin axis, the 
electrons will collapse to similar spins, with the spin state determined by the 
instant of measurement if the entangled pair is described by the Bell state 
$|\psi_{1}\rangle$ in Figure ~\ref{fig:fig3}. Likewise, it will collapse to 
opposite spins if the entangled pair is described by $|\psi_{2}\rangle$ in 
Figure ~\ref{fig:fig3}.

\begin{figure}[ht]
\begin{center}
\includegraphics[width=80mm,height=50mm]{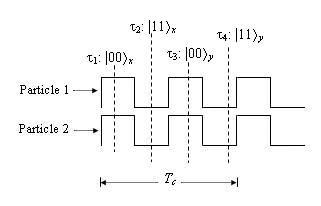}
\caption{Two-particle entangled system with non-commutating observables.}\label{fig:fig4}
\end{center}
\end{figure}

(2) \textit{Non-commutating observables}. Consider an entangled pair of 
electrons where the electrons have spin components along two axes, say, $x$-axis 
and $y$-axis (see Figure 4). Note that at any given instant the spin of both 
electrons will be along only one of the axes. Further, at any instant, such as 
$\tau_{1}$, $\tau_{2}$, $\tau_{3}$, and $\tau_{4}$ shown in Figure ~\ref{fig:fig4}, 
the two electrons can be in only one of the four states: $|00\rangle_{x}$, 
$|11\rangle_{x}$, $|00\rangle_{y}$, and $|11\rangle_{y}$, respectively. The 
suffixes $x$, $y$ represent the $x$, $y$ components, respectively, of 
$|00\rangle$ and $|11\rangle$.

Now, if a measurement is made and the system collapses, say, to the $x$-component, 
then a subsequent measurement along the $y$-axis will return a null result.

In the more general case of a system of $n$-particles, if a combined measurement 
is made on $m \leq n$ of those particles then those $m$-particles will collapse to 
one of their possible group states (the actual number of states at any given 
instant may vary from $1$ to $2^{m}$) on measurement while the remaining $n - m$ 
particles will assume states which are consistent with the collapsed state of the 
$m$-particles. 

\subsection*{Quantum adder}
\begin{figure}[ht]
\begin{center}
\includegraphics[width=130mm,height=55mm]{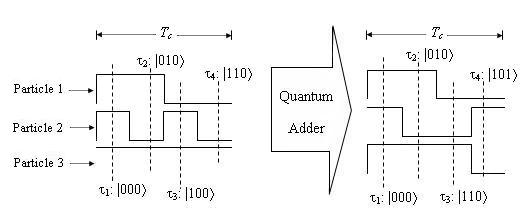}
\caption{Quantum adder input and output states; $|\psi_{1}\rangle$ (left), 
         $|\psi_{2}\rangle$ (right).}\label{fig:fig5}
\end{center}
\end{figure}

The quantum adder operation can be explained by our interpretation in a consistent 
manner. The initial input state of the required three particle system, where each 
particle represents a qubit, is given by $|\psi_{0}\rangle = |000\rangle$. We now 
apply the Hadamard gate to the first two qubits to create the four possible inputs 
for the addition operation. Thus we have 
\begin{equation*}
  |\psi_{1}\rangle = (|000\rangle + |010\rangle + |100\rangle + |110\rangle)/2.
\end{equation*}
To carry out the add operation we apply the Toffoli gate to the three qubits with 
the third qubit as target, followed by the $C_{\textit{not}}$ gate to the first two 
qubits with the second qubit as target to get
\begin{equation*}
  |\psi_{2}\rangle = (|000\rangle + |010\rangle + |110\rangle + |101\rangle)/2,
\end{equation*}
where the second qubit is the sum and the third qubit is the carry bit. Note that 
the carry bit in the adder is the result of an AND operation. The carry and AND 
are really the same thing. The sum bit comes from an XOR gate (that is, the 
C$_{\textit{not}}$ operation). Figure ~\ref{fig:fig5} captures the four possible 
eigenstates represented by $|\psi_{1}\rangle$ and $|\psi_{2}\rangle$ at the 
instants $\tau_{1}$, $\tau_{2}$, $\tau_{3}$, and $\tau_{4}$.

\subsection*{Teleporting a qubit of an unknown state}
\begin{figure}[ht]
\begin{center}
\includegraphics[width=85mm,height=63mm]{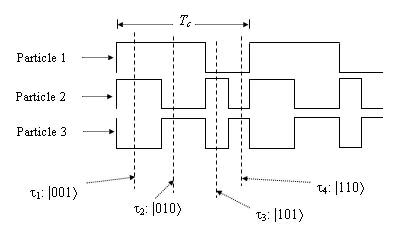}
\caption{Initial state $|\psi_{0}\rangle$ of the teleportation system.}\label{fig:fig6}
\end{center}
\end{figure}

\begin{figure}[ht]
\begin{center}
\includegraphics[width=85mm,height=63mm]{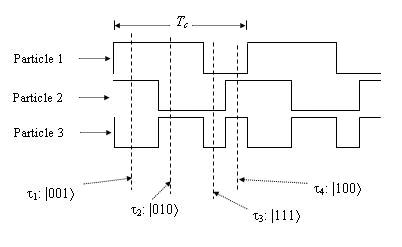}
\caption{State $|\psi_{1}\rangle$ of the teleportation system after $C_{\textit{not}}$
         operation.}\label{fig:fig7}
\end{center}
\end{figure}

\begin{figure}[ht]
\begin{center}
\includegraphics[width=90mm,height=60mm]{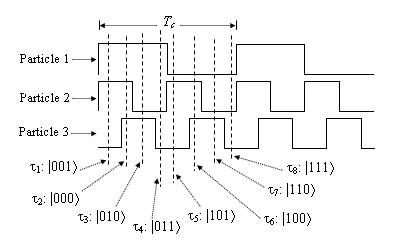}
\caption{State $|\psi_{2}\rangle$ of the teleportation system after Hadamard 
         operation.}\label{fig:fig8}
\end{center}
\end{figure}

In our final example we show that the teleportation method of Bennett \textit{et 
al} $\cite{R15}$ is also correctly explained by our interpretation. Here Alice 
wishes to teleport a qubit, labelled by subscript 1, of unknown state 
$|\phi\rangle = \alpha |0_{1}\rangle + \beta |1_{1}\rangle$, to Bob. In addition, 
there is an entangled pair of auxiliary qubits designated by subscripts 2 and 
3 in the state $|\chi\rangle = (|0_{2}1_{3}\rangle - |1_{2}0_{3}\rangle)/\sqrt{2}$. 
Alice holds the qubit with subscript 2 in addition to the one with subscript 1
while Bob holds the qubit with subscript 3. Thus the initial state of the three 
qubit system is (see Figure ~\ref{fig:fig6} where the qubit subscripts (in this 
and subsequent Figures ~\ref{fig:fig7} and ~\ref{fig:fig8}) have been omitted 
since they can be inferred from their position in the state $|\ldots\rangle$) 
given by
\begin{equation*}
  |\phi\chi\rangle = |\psi_{0}\rangle = 
                     [\alpha |0_{1}\rangle (|0_{2}1_{3}\rangle - |1_{2}0_{3}\rangle) + 
                      \beta |1_{1}\rangle (|0_{2}1_{3}\rangle - |1_{2}0_{3}\rangle)]/\sqrt{2}  
\end{equation*}
Alice now applies the $C_{\textit{not}}$ gate (with qubit 2 as the target) to the 
qubits held by her. This changes the state of the three qubit system to (see 
Figure ~\ref{fig:fig7})
\begin{equation*}
  |\psi_{1}\rangle = [\alpha |0_{1}\rangle (|0_{2}1_{3}\rangle - |1_{2}0_{3}\rangle) + 
                      \beta |1_{1}\rangle (|1_{2}1_{3}\rangle - |0_{2}0_{3}\rangle)]/\sqrt{2}  
\end{equation*}
Next, Alice applies the Hadamard gate to qubit 1 which puts the three qubit 
system in the state (see Figure ~\ref{fig:fig8})
\begin{align*}
  |\psi_{2}\rangle = [&|0_{1}0_{2}\rangle (\alpha|1_{3}\rangle - \beta|0_{3}\rangle) -
                      |0_{1}1_{2}\rangle (\alpha|0_{3}\rangle - \beta|1_{3}\rangle) + \\
                      &|1_{1}0_{2}\rangle (\alpha|1_{3}\rangle + \beta|0_{3}\rangle) -
                      |1_{1}1_{2}\rangle (\alpha|0_{3}\rangle + \beta|1_{3}\rangle)]/2
\end{align*}

\clearpage
Finally, Alice makes a ``combined'' measurement on the two qubits she holds. Such 
a measurement gives access to some combined (or global) information on both qubits, 
but none on a single qubit, that is, no distinction between the two qubits can be 
established. Her measurement will lead the pair to collapse to one of the four 
possible states $|0_{1}0_{2}\rangle$, $|0_{1}1_{2}\rangle$, $|1_{1}0_{2}\rangle$, 
or $|1_{1}1_{2}\rangle$, while the third qubit, correspondingly, will immediately 
collapse to the state $\alpha|1_{3}\rangle - \beta|0_{3}\rangle$, 
$\alpha|0_{3}\rangle - \beta|1_{3}\rangle$, $\alpha|1_{3}\rangle + \beta|0_{3}\rangle$
or $\alpha|0_{3}\rangle + \beta|1_{3}\rangle$, respectively, since it is also 
entangled with qubits 1 and 2. Table ~\ref{tab:tab1} shows the measurement result 
Alice will get depending upon the instant the measurement actually occurred, along 
with the post-measurement state of qubit 3 held by Bob. 

\begin{table}[ht]
\begin{center}
\caption{Measurement outcomes in the teleportation algorithm.}
\centering
\begin{tabular}{| c | c | c | c | c |}
\hline \hline
Measurement & State of qubits 1 \& 2 & State of qubit 3  & \multicolumn{2}{c}{Decoder to bring} \vline\\
instant     & after Alice's          &  after Alice's    & \multicolumn{2}{c}{qubit 3 to state $\phi\rangle$} \vline\\
            &  measurement           &  measurement      & \multicolumn{2}{c}{}\vline \\
\hline
 & & & & $|0_{3}\rangle \rightarrow -|1_{3}\rangle$ \\ 
\raisebox{1.5ex}{$\tau_{1}$,$\tau_{2}$} & \raisebox{1.5ex}{$|00\rangle$} & 
\raisebox{1.5ex}{$\alpha|1\rangle-\beta|0\rangle$} & \raisebox{1.5ex}{Y} & $|1_{3}\rangle \rightarrow |0_{3}\rangle$ \\
\hline
 & & & & $|0_{3}\rangle \rightarrow |0_{3}\rangle$ \\ 
\raisebox{1.5ex}{$\tau_{3}$,$\tau_{4}$} & \raisebox{1.5ex}{$|01\rangle$} & 
\raisebox{1.5ex}{$\alpha|0\rangle-\beta|1\rangle$} & \raisebox{1.5ex}{Z} & $|1_{3}\rangle \rightarrow -|1_{3}\rangle$ \\
\hline
 & & & & $|0_{3} \rightarrow |1_{3}\rangle$ \\ 
\raisebox{1.5ex}{$\tau_{5}$,$\tau_{6}$} & \raisebox{1.5ex}{$|10\rangle$} & 
\raisebox{1.5ex}{$\alpha|1\rangle+\beta|0\rangle$} & \raisebox{1.5ex}{X} & $|1_{3}\rangle \rightarrow |0_{3}\rangle$ \\
\hline
 & & & & $|0_{3} \rightarrow |0_{3}\rangle$ \\ 
\raisebox{1.5ex}{$\tau_{7}$,$\tau_{8}$} & \raisebox{1.5ex}{$|11\rangle$} & 
\raisebox{1.5ex}{$\alpha|0\rangle+\beta|1\rangle$} & \raisebox{1.5ex}{I} & $|1_{3}\rangle \rightarrow |1_{3}\rangle$ \\
\hline
\end{tabular}
\label{tab:tab1}
\end{center}
\end{table}
Alice communicates the classical result of her ``combined'' measurement 
($|0_{1}0_{2}\rangle$, $|0_{1}1_{2}\rangle$, $|1_{1}0_{2}\rangle$, or 
$|1_{1}1_{2}\rangle$) to Bob (using classical means such as telephone, email, 
etc.). Bob then uses the decoder (a unitary transformation) listed in Table 
~\ref{tab:tab1} corresponding to the state of qubits 1 \& 2 conveyed to him 
by Alice to bring his qubit to state $|\phi\rangle = \alpha |0_{3}\rangle + 
\beta |1_{3}\rangle$.


\section*{Conclusion}
We have presented a new interpretation of the terms superposition, entanglement, 
and measurement that appear in quantum mechanics. In this interpretation the 
structure of the wave function at the sub-Planck scale has a deterministic cyclic 
structure. Our attempt has been to provide minimal structure at the sub-Planck 
level for it to match observations. This structure is independent of the 
Schr\"{o}dinger wave equation. Each cycle comprises a sequential succession 
of the eigenstates that comprise a given wave function.

In our interpretation, superposed states appear as time-sliced in a cyclic manner 
such that the time spent by an eigenstate in a cycle is related to the complex 
amplitudes appearing in the wave function. Entangled states binding two or more 
particles appear in our interpretation as the synchronization of the sub-Planck 
level oscillation of the participating particles.

The probabilistic aspect of quantum mechanics is interpreted by hypothesizing a 
measurement mechanism which acts instantaneously but the instant of measurement 
is chosen randomly by the classical measuring apparatus over a small but finite 
interval from the time the measurement apparatus is activated. At the instant the 
measurement is made, the wave function irrevocably collapses to a new state and 
continues from thereon in that state till changed by a unitary operation or a new 
measurement.

We hope our interpretation makes quantum theory more intelligible and intuitively 
acceptable to the human mind. As a working strategy, we have no issue with the 
Copenhagen interpretation, but at a psychological level humans are more 
comfortable with an interpretation they can intuitively relate to. Our 
interpretation creates a hypothetical mechanism at the sub-Planck level, which 
supposedly drives quantum mechanical phenomenon. We believe our interpretation 
has value because it appears to be compatible with measurements at the classical 
level. We are not physically equipped to perceive the quantum world and our 
measuring devices are classical. We can therefore only speculate what Nature is 
like.

\bibliography{main}

\begin{thebibliography}{10}

\bibitem{R1}
Nielsen, M. A., and Chuang, I. L.,
\newblock {\em Quantum Computation and Quantum Information},
\newblock Cambridge University Press, Cambridge, 2000.
\newblock [For errata see: \url{http://www.squint.org/qci/}].

\bibitem{R2}
Albert, D.,
\newblock {\em Quantum Mechanics and Experience},
\newblock Harvard University Press, Cambridge, MA, 1992. p. 79. 

\bibitem{R3}
Lalo\"{e}, F.,
\newblock Do we really understand quantum mechanics? Strange correlations, paradoxes, and theorems.
\newblock {\em Am. J. Phys.} \textbf{69}(6), June 2001, pp. 655-701.
\newblock [For an updated version see \url{http://hal.archives-ouvertes.fr/docs/00/02/75/08/PDF/mq-anglais.pdf}].

\bibitem{R4}
Podolsky, B., Einstein, A., and Rosen, N.,
\newblock Can quantum-mechanical description of physical reality be considered
  complete?
\newblock {\em Phys. Rev.} \textbf{47},777-780 (1935).

\bibitem{R5}
Bell, J. S.,
\newblock On the Einstein-Podolsky-Rosen paradox.
\newblock {\em Physics}, \textbf{1}, 195-200, 1964.
\newblock Reprinted in J. S. Bell, {\em Speakable and Unspeakable in Quantum
  Mechanics}, Cambridge University Press, Cambridge, 1987.

\bibitem{R6}
Aspect, A., Dalibard, J., and Roger, G.,
\newblock Experimental test of Bell's inequalities using time-varying analysers.
\newblock {\em Phys. Rev. Lett.} \textbf{49}, (1982), pp. 1804-1807.
\newblock See also Aspect, A., Testing Bell's inequalities, Europhys. News,
  \textbf{22}, 73-75 (1991).

\bibitem{R7}
Hofstadter, D., 
\newblock {\em G\"{o}del, Escher, Bach.},
\newblock Vintage Books, New York, 
\newblock 1989.

\bibitem{R8}
Jammer, M.,
\newblock {\em The Philosophy of Quantum Mechanics},
\newblock Wiley \& Sons, New York, 1974, p. 87.
\newblock (Quotation as reproduced in Al-Kahlili, J., {\em Quantum}, Weidenfeld \&
  Nicolson, London, 2003, p. 134.)

\bibitem{R9}
Born, M.,
\newblock Zur Quantenmechanik der Sto\ss vorg\"{q}nge.
\newblock {\em Z. Phys.}, \textbf{37}, 863-867, July (1926).
\newblock English translation in Ludwig, G., ed., 1968, {\em Wave Mechanics},
  Pergamon Press, Oxford, p. 206.

\bibitem{R10}
Seife, C.,
\newblock Do deeper principles underlie quantum uncertainty and nonlocality.
\newblock {\em Science}, \textbf{309}, 1 July 2005. p. 98.

\bibitem{R11}
Al-Kahlili, J.,
\newblock {\em Quantum},
\newblock Weidenfeld \& Nicolson, 2003.

\bibitem{R12}
Everett, H.,
\newblock ``Relative State" formulation of quantum mechanics.
\newblock {\em Reviews of Modern Physics}, Vol. \textbf{29}, No. 3, July 1957.
\newblock \url{http://www.univer.omsk.su/omsk/Sci/Everett/paper1957.html}.

\bibitem{R13}
Byrne, P.,
\newblock The many worlds of Hugh Everett.
\newblock {\em Scientific American}, December 2007,
\newblock \url{http://www.scientificamerican.com/article.cfm?id=hugh-everett-biography}.

\bibitem{R14}
Bohm, D., and Hiley, B. J.,
\newblock {\em The Undivided Universe}, Routledge, London, 1993, Chapter 3.
\newblock The original paper is: Bohm, D., An interpretation in terms of
  hidden variables, Phys. Rev. \textbf{85}, 166-193, 1952.

\bibitem{R15}
Bennett, C. H., Brassard, G., Cr\'{e}peau, C., Jozsa, R., Peres, A., and Wootters, W.,
\newblock Teleporting an unknown quantum state via dual classical and EPR channels.
\newblock {\em Phys. Rev. Lett.} \textbf{70}, 1895-1899, 1993.
\newblock \url{http://www.research.ibm.com/quantuminfo/teleportation/teleportation.html}.

\end{thebibliography}
\bibliographystyle{plain}
\end{document}